\documentclass[aps,pra,superscriptaddress,12pt]{revtex4}
\usepackage[intlimits]{amsmath}
\usepackage{amsfonts}
\usepackage{psfrag}
\usepackage{subfigure}
\usepackage[usenames]{color}
\usepackage{ifpdf}

\ifpdf
\usepackage{epstopdf}
\usepackage[pdftex]{hyperref}
\else
\usepackage[hypertex,ps2pdf]{hyperref}
\fi

\advance\voffset by -1cm
\advance\hoffset by -0.5cm
\newcommand\beq            {\begin{equation}}
\newcommand\bea           {\begin{equation}\begin{array}l\displaystyle}
\newcommand\eeq            {\end{equation}}
\newcommand\bes           {\begin{subequations}}
\newcommand\esu           {\end{subequations}}

\begin{document}


\author{L. Lepori}
\affiliation{SISSA and INFN, Sezione di Trieste, via Bonomea 265, 
I-34136 Trieste, Italy}

\author{G. Mussardo}
\affiliation{SISSA and INFN, Sezione di Trieste, via Bonomea 265, 
I-34136 Trieste, Italy}
\affiliation{International Centre for Theoretical Physics (ICTP), 
Strada Costiera 11, I-34151, Trieste, Italy}

\author{A. Trombettoni}
\affiliation{SISSA and INFN, Sezione di Trieste, via Bonomea 265, 
I-34136 Trieste, Italy}

\title{$(3+1)$ Massive Dirac Fermions with Ultracold Atoms \\ 
in Optical Lattices}


\begin{abstract}
\noindent 
We propose the experimental realization of $(3+1)$ relativistic Dirac fermions 
using ultracold atoms in a rotating optical lattice or, alternatively, 
in a synthetic magnetic field. 
This approach has the advantage  
to give mass to the Dirac fermions by coupling the ultracold 
atoms to a Bragg pulse. A dimensional crossover from $(3+1)$ to $(2+1)$ 
Dirac fermions can be obtained by varying the anisotropy of the lattice. 
We also discuss under which conditions the interatomic potentials 
give rise to relativistically invariant interactions among the Dirac fermions.
\end{abstract}
\maketitle

\newpage

For their high level of control, trapped ultracold 
atoms are ideal systems for simulating in a tunable way 
strongly interacting models \cite{bloch08}. 
A well-known example is the experimental realization 
of interacting lattice Hamiltonians: for bosonic gases, the Mott-superfluid 
transitions has been both detected \cite{greiner02} and investigated in a variety of interesting situations, including low dimensional and disordered set-ups \cite{bloch08}; 
for fermionic gases, the recent studies \cite{schneider08,jordens08} of metallic and insulating phases of a two-species mixture in a $3D$ optical lattice have opened 
the way to experimentally investigating the rich phase diagram of the Fermi-Hubbard model.

The examples mentioned above refer to the ability of cold atom systems to simulate non-relativistic Hamiltonians but a fascinating new challenge of the field is the tunable experimental realization of relativistic systems which are relevant to 
high energy physics and quantum gauge theories \cite{conf}. It is worth mentioning, for instance, the simulation of the properties of graphene \cite{castro09}, i.e. $(2+1)$ relativistic Dirac fermions, obtained by  
using ultracold fermions in honeycomb lattices 
\cite{duan07,guineas,wu08,lee09}.
Other recent proposals to realize massless $(2+1)$ Dirac fermions consist of ultracold 
fermions on a square lattice coupled with properly chosen Rabi fields 
\cite{hou09}, interacting bosons in a two-dimensional lattice produced 
by a bichromatic light-shift potential with an additional effective 
magnetic field \cite{lim08} and bosons with internal energy levels in a tripod configuration \cite{Juzeliunas}.

It is then highly interesting to see whether it is possible to go beyond the (2+1) case and simulate relativistic (3+1) Dirac fermions. We are concerned, in particular, with the possibility to make them massive and also interacting, possibly in a Lorentz invariant way.
Mixtures of two ultracold fermionic species (and recently of three species \cite{wille08,ottenstein08}) may also be useful for the experimental realization of Dirac fermions with internal degrees of freedom. New developments in this direction could open the way to simulate, by cold atom systems, Kogut-Susskind staggered lattice fermions \cite{kogut75,susskind77}  or more general elementary particle theories. In perspective, this development could permit to study in a controllable experimental set-up part of the phase diagram of QCD \cite{Rajagopal}.

The aim of this paper is to discuss an experimental scheme  to realize  $(3+1)$ massive Dirac fermions (with a mass eventually time-dependent) using ultracold atomic fermions, a set-up which makes possible to control interactions through Feshbach resonances \cite{bloch08} and to realize mixtures of different internal states. A method of 
simulating the Dirac equation in $(3+1)$ dimensions for a free spin-$1/2$ 
particle in a single trapped ion was presented in \cite{Lamata}, where the 
transition from massless to massive fermions was also studied. 
Here we propose instead to use non-relativistic polarized ultracold fermions 
in a rotating cubic optical lattice with tight-binding Hamiltonian 
\beq
H= - t \sum_{\langle i, j \rangle}  \left( c_{i}^{\dag} e^{-i A_{ij} } c_{j} + h.c. \right) \, ,
\label{tbham}
\eeq
where $c_{i}^{\dag}$ creates an atom in the $i$-th well of the lattice, $t$ is the tunneling parameter (assumed for the moment equal along the three axes $x$, $y$ and $z$) and the sum is on nearest-neighbours wells. The lattice (created with three counterpropagating laser beams) is assumed to be rotating with angular 
velocity $\vec{\omega}$ so that the electrically neutral atoms feel an effective magnetic field: with the minimal substitution $-i \hbar \vec{\nabla} \to 
-i \hbar \vec{\nabla}-m\vec{A}$ (where $m$ is the mass of the atoms and $\vec{A}=\vec{\omega} \times \vec{r}$ is the analog of the magnetic vector potential) we have the Hamiltonian (\ref{tbham}), with 
$A_{ij}=(m/\hbar) \int_i^j \vec{A} \cdot d\vec{l}$. Rotating lattices have been efficiently realized quite recently employing four intersecting laser beams manipulated with acousto-optical deflectors \cite{williams10}). Alternatively, one could also end up in the Hamiltonian (\ref{tbham}) using, on a cubic lattice, fermions subjected to a synthetic magnetic field obtained by spatially dependent optical coupling between internal states of the atoms \cite{lin09}. 

Before studying the spectrum of the Hamiltonian (\ref{tbham}), let's briefly comment on the reason of its choice: one may wonder, in fact, if a simpler Hamiltonian -- without a magnetic field -- of the form $H= - t \sum_{i, j}  c_{i}^{\dag} B_{ij} c_{j}$ ($B_{ij}=1$ if $i$ and $j$ are 
nearest-neighbours and $0$ otherwise) is able to simulate $(3+1)$ Dirac fermions. This is equivalent to ask whether it is possible to realize, with a suitable choice of $B_{ij}$, a semi-metal such that the bands touch at {\em isolated} points. 
In \cite{abrikosov70} the symmetries groups which lead to a spectrum without Fermi surface and energy gap were classified: although this result can be used to exclude certain classes of $B$'s matrices, it does not help however to identify  
the tight-binding Hamiltonians which could have the desired spectral properties. By direct inspection, we checked that the $3D$ Bravais lattices with a single atom per cell and only nearest-neighbour hoppings does not give band touching in isolated points at zero energy. It is for this reason that we focus our attention on the realization of Dirac fermions using an artificial uniform 
magnetic field $\vec{B} = {\rm rot}\vec{A} = \pi \phi_0
(1,1,1)$, corresponding to $\omega=(h/8ma^2) (1,1,1)$ 
($a$ is the lattice spacing and $\phi_0=\hbar/2ma^2$). 
With $a \sim 2 \mu m$, one gets 
for $K$ atoms a rotation frequency 
$\nu \equiv \vert 
\vec{\omega} \vert / 2 \pi\sim 
200 Hz$. 
From now on we set $\phi_0=1$ and $a=1$.

The magnetic field $\vec{B} = \pi (1,1,1)$ induces a $\pi$-flux on every square face 
(see the schematic plot in Fig.1): 
to diagonalize Hamiltonian (\ref{tbham}) is mostly convenient to use the gauge 
$\vec{A}= \pi (0, x-y, y-x)$, similarly to 
\cite{Hasegawa}. The quasimomenta $\vec{k}$ take the values 
in the magnetic Brillouin zone \cite{landau9}, given by 
$-\pi/2 < k_{x,y} < \pi/ 2$, 
$- \pi < k_z < \pi$ \cite{nota1}. 

Since the lattice sites can be divided in two inequivalent sets, say $A$ and $B$ 
(see Fig.1), we can write the Fourier transforms as  
$c_{\Gamma}(\vec{k})=\sum_{j \in \Gamma } c_j e^{i \vec{k} \cdot \vec{j}}$ (where 
$\Gamma=A,B$) and plugging in eqn. (\ref{tbham}), 
we get the spectrum \cite{Hasegawa,Zou}:
\beq
E(\vec{k})= \pm 2 \, t \, \sqrt{cos^2 k_x +cos^2 k_y +cos^2 k_z} \, .
\label{energy}
\eeq
An energy spectrum like (\ref{energy}) was obtained for $PbTe$-type narrow-gap semiconductors with antiphase boundaries \cite{fradkin86}; a model having this spectrum has been recently used in \cite{maraner08}, where it was shown 
that a suitable distortion of tunneling couplings in fermionic lattices can introduce a scalar and a Yang-Mills field.

\begin{figure}[b]
\centerline{\scalebox{0.35}{\includegraphics{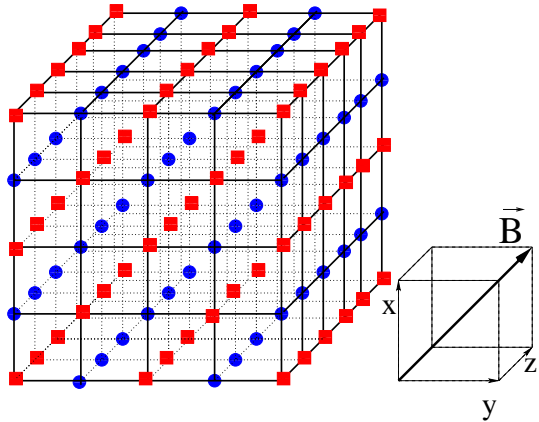}}}
\caption{Sets $A$ (red squares) and $B$ (blue circles) 
on the cubic lattice (right: 
artificial magnetic field $\vec{B}$).}
\end{figure}

For half-filling the Fermi energy is zero and there is a vanishing gap between valence ($E<0$) and conducting bands ($E>0$) at the {\em isolated} Dirac points 
$\vec{k}= \pm \frac{\pi}{2} (\pm 1, \pm 1, \pm 1)$. A pair of inequivalent 
Dirac points is given by $\vec{k}_R = \frac{ \pi }{2}(1,1,1)$ and 
$\vec{k}_L = - \frac{ \pi }{2}(1,1,1)$. 
Expanding the energy around these 
Dirac points we have  $E(\vec{k}_{L/R}+\vec{q})/\hbar \approx v_F  |\vec{q}|$, where the Fermi velocity is given by $v_F=2ta/\hbar$. Close to these zero-gap points,  the quasiparticles behave as massless $(3+1)$ Dirac fermions of both the chiralities \cite{Nielsen1} and the linearized form of the Hamiltonian (\ref{tbham}) becomes, in the continuum limit, the 3-d Dirac Hamiltonian
\beq
H= -2i t \int d\vec{r} \,  
\left(\psi_{R}^\dag \vec{\sigma} \cdot \vec{\nabla} 
\psi_{R} - \psi_{L}^\dag \vec{\sigma} 
\cdot \vec{\nabla} \psi_{L}\right)
\label{ham}
\eeq 
where $\vec{\sigma}=(\sigma_x,\sigma_y,\sigma_z)$ are the Pauli matrices 
and the two-components spinors 
$\psi_{L}(\vec{r})$, $\psi_{R}(\vec{r})$ 
are respectively the Fourier transforms of 
\beq 
\psi_{R}(\vec{k})=  \left(\begin{array}{c} c_A(\vec{k} +\vec{k_R}) \\
c_B (\vec{k} + \vec{k_R}) \end{array} \right) ; 
\psi_{L}(\vec{k})=  \left(\begin{array}{c} c_A(\vec{k} + \vec{k_L}) \\  
c_B (\vec{k} + \vec{k_L}) \end{array}\right) .
\label{spinors}
\eeq

In experimental realizations, the magnetic field $\vec{B} = \pi (1,1,1)$ may be subjected to some fluctuations which change the magnetic flux per plaquette, $2\pi \Phi$, around the value $\Phi = 1/2$. In the thermodynamical limit $L \to \infty$, where $N=L^3$ is the number of sites of the cubic lattice, these fluctuations are expected to influence the Dirac cones because, when the flux on a plaquette is different from a rational number $\frac{p}{q}$, the usual Bloch functions are no longer a faithful representation of the translation group and therefore the energy spectrum assumes a fractal structure \cite{Hofstadter,koshino,BDG}. Note, however, that: (a) experiments with trapped ultracold atoms are done with finite number of atoms ($N \sim 10^3-10^5$); (b) for finite $L$, the spectrum is not sensibly affected by fluctuations of $\Phi$ which are much smaller than $1/L$. To clarify this point, consider two close rational values of the flux, 
say $\frac{1}{2}$ and $\frac{51}{100}$: in the second case one has $q=100$ 
inequivalent vertices (and $q$ sub-bands), while there are only $q=2$ in the first case. 
However, if $L \ll 1/\delta \Phi$, the $q$ sub-bands are gathered in two groups, each of them almost degenerate. Hence, for the two values of the flux we have two physical situations which, for realistic numbers $L$ and $\delta \Phi \ll 1$, 
are practically indistinguishable. This can be explicitly checked by numerically diagonalizing the Hamiltonian (\ref{tbham}): e.g., even for a rather small size ($L=16$) 
with open boundary conditions and a fairly large value of $\delta \Phi$ ($\sim 5\%$ of $\Phi=1/2$), the spectral density is not sensibly affected and it is in reasonable 
agreement with the one computed from (\ref{energy}).

Let now expose the ultracold atomic gas to a Bragg pulse (see for example \cite{blakie02,bragg}). For deep optical lattices, the Hamiltonian (\ref{tbham}) acquires a new term of the form 
\beq
H_B = V_0 \left( \sum_{j} c_j^{\dag} c_j \, e^{i \vec{k}_{Bragg} \cdot \vec{j} } 
e^{-i \omega t}  \, + h.c. \right) 
\label{bragham}
\eeq
where the sum runs on the lattice sites while $\vec{k}_{Bragg}$ and $\omega$ are the differences between the wave-vectors and the frequencies of the used lasers. The Bragg term (\ref{bragham}) gives rise to a mass of the Dirac fermions: choosing $\vec{k}_{Bragg} = \vec{k}_L-\vec{k}_R=(\pi,\pi,\pi)$, the 
quasiparticles around the Dirac point $\vec{k}_L$ ($\vec{k}_R$) are 
transferred close to the Dirac point $\vec{k}_R$ ($\vec{k}_L$) 
inverting the chirality. This is equivalent to add a mass term to the Dirac Hamiltonian (\ref{ham}) 
\beq 
\frac{V_0}{2} \, cos(\omega t) (\psi_L^{\dag} \psi_R \, + \, \psi_R^{\dag} \psi_L ) = \frac{V_0}{2} \, cos(\omega t) \, \bar{\psi} \psi 
\label{mass}
\eeq
where $\psi=\left(  \begin{array}{c} \psi_{R}\\ \psi_{L} \end{array} \right)$ 
and $\bar{\psi}=\psi^{\dag } 
\left( \begin{array}{cc}  \mathbf{0} & \mathbf{1} 
\\ \mathbf{1} & \mathbf{0}  \end{array} 
\right) $ ($\mathbf{0}$ and $\mathbf{1}$ are the 
$2 \times 2$ zero and identity matrices).
Notice that above we exploited the periodicity of the magnetic dual lattice.
When the frequency difference vanishes, $\omega=0$, one has a time-independent Dirac mass while, keeping $\vec{k}_{Bragg}$ fixed, but changing randomly the 
intensity of the two lasers, 
one has also the interesting possibility to realize a Dirac fermion with random
mass. 

The term (\ref{bragham}) is actually a particular case of a more general situation: starting from the Hamiltonian 
$H= -t \sum_{\langle i, j \rangle}  c_{j}^{\dag} \, e^{-i (A_{ij} + {\cal A}_{ij}) } c_{i} 
+\sum_j A_0 (j) \,  c^{\dag}_j c_j$ and 
repeating almost unchanged the calculation that led to (\ref{ham}), one gets 
in the continuum limit the Dirac Hamiltonian in an e.m. field
\beq
\begin{array}{c}
H=-2it \int d\vec{r} \,  \left( \psi_{R}^\dag(\vec{r}) \, 
\vec{\sigma} \cdot \vec{D} \, \psi_{R}(\vec{r}) -
\psi_{L}^\dag(\vec{r}) \, \vec{\sigma} \cdot \vec{D} \, 
\psi_{L}(\vec{r}) \right) \\
+ \int d\vec{r} \, A_0 \left( \psi_{R}^\dag(\vec{r})  \psi_{R}(\vec{r})  +  
\psi_{L}^\dag(\vec{r}) \psi_{L}(\vec{r}) \right)
\end{array}
\label{ham.em}
\eeq
where the perturbations ${\cal A}_{ij}$ and $A_0$ 
are slowly varying in space and time and $\vec{D}= \vec{\nabla} + \vec{\cal A}$. 

So far we have considered equal hopping parameters along the $x$, $y$ 
and $z-$axes but, since the tunneling rates depend on the power of the lasers, one can easily realize different hopping parameters $t_x , t_y, t_z $. 
The energy spectrum in this case is
\beq
E(\vec{k})= \pm 2 \,  \sqrt{t_x^2 \, cos^2 k_x + t_y^2 \, cos^2 k_y + t_z^2 \, cos^2 k_z} \, .
\label{energy_anis}
\eeq
and the isolated Dirac points are therefore unaffected by the anisotropy of the hopping parameters. The energy spectrum can be also derived including next-nearest-neighbour 
hopping rates and if they are small with respect to $t$, 
as it happens in graphene \cite{castro09}, 
the low-energy dynamics is still well described by Dirac fermions.  

Using these anisotropic hopping parameters, 
we can easily realize the crossover 
from $(3+1)$ to $(2+1)$ Dirac fermions: to this aim, it is sufficient to lowering an hopping parameter to zero (say $t_z$, amounting to increase the power of the laser along $z$) while keeping fixed the magnetic field. When $ t_z \to 0 $, for the fermions we have 
$\psi_R (\vec{k} + \vec{k_R}) 
\to \sigma_x \psi_1 (\vec{k} + \vec{k}_R^{\prime})$ and 
$\psi_L (\vec{k} + \vec{k_L})  \to \sigma_x \sigma_z  \psi_2 (\vec{k} + 
\vec{k}_L^{\prime})$, 
with $\vec{k}_R^{\prime} = (\frac{\pi}{2}, -\frac{\pi}{2}, -\frac{\pi}{2})$ 
and $\vec{k}_L^{\prime} = (-\frac{\pi}{2}, \frac{\pi}{2}, -\frac{\pi}{2})$,
clearly equivalent to $\vec{q}_R$ and $\vec{q}_L$.  The Hamiltonian, on the other hand, becomes a purely two-dimensional one  
\begin{eqnarray*}
H^{2D}= 2t \int d\vec{p} \, \left( \bar{\psi}_{1}^{2D}(\vec{p}) \, \vec{\alpha} \cdot \vec{p} \; \psi_{1}^{2D}(\vec{p})   \right.\\
\,\,\,\,\,\,\,+ \,\left.\bar{\psi}_{2}^{2D}(\vec{p}) \, \vec{\alpha} \cdot \vec{p} \; \psi_{2}^{2D}(\vec{p}) \right) \, ,
\label{ham3-2D}
\end{eqnarray*}
where $\vec{p}=(k_x,k_y)$ is the quasimomentum on the $x-y$ plane, 
$\vec{\alpha} = i (\sigma_x , \sigma_y)$, $\gamma_0 = \sigma_{z}$ and 
\[
\psi_{1,2}^{2D}(\vec{p}) = \lim_{k_z \to 0} e^{i \frac{\pi}{4} \, \sigma_z} \, \psi_{1,2} \left( \vec{k} \pm \left(\frac{\pi}{2} ,  \pm  \frac{ \pi}{2},   \frac{ \pi}{2} \right) \right) \, .
\]
This is nothing else that the Hamiltonian for $(2+1)$ Dirac fermions 
obtained in \cite{Semenoff}. Hence, in the $2D$ limit,  
we obtain directly a pair of $(2+1)$ massless fermions, 
as expected \cite{Affleck}. One can also show that, 
in the limit $t_z \to 0 $,  
the Bragg term (\ref{bragham}) gives mass also to the $(2+1)$ Dirac fermions.

Notice that, in the discussion above, 
one can consider two (or eventually more) 
fermionic species: they can be either different hyperfine levels of the same 
fermionic species or different species of a mixture 
(e.g., a $Li$-$K$ mixture). Experiments with collisionally stable mixtures 
of two \cite{bloch08} and also three \cite{wille08,ottenstein08}) 
fermionic species has been recently reported. The low-energy 
Hamiltonian will be then simply the sum of free Hamiltonians 
of the type (\ref{ham}): in this scheme, the 
mass of different Dirac fermions obtained by Bragg pulses
can be in principle different.

Let now finally consider two-body interactions among ultracold fermions 
of the form  
\beq
\sum_{i,j} \, U_{i,j} \,  
c_{i}^{\dag} c_{i} c_{j}^{\dag} c_{j} \,:
\label{interaz}
\eeq
eqn (\ref{interaz}) describes general non-local interactions among atoms 
of the same species, as it may be realized in 
$p$-wave channels \cite{gurarie07}. We assume 
for simplicity that $U_{ij}$ is a function only of 
$\vec{r}=\vec{i}-\vec{j}$. It is well known that the interatomic term (\ref{interaz}) 
does not generally give rise to a Lorentz invariant interaction 
among the Dirac spinors. 
In our case we point out that, to have Lorentz invariance, 
a necessary condition is      
\beq
\tilde{U}(0)= 0 \,,
\label{trasf} 
\eeq 
where $\tilde{U}(\vec{k})$ denotes the Fourier transform of $U(\vec{r})$. This condition  implies in fact the correct inversion of the chirality of the bilinears entering the terms like $\psi^\dag \psi^\dag \psi \psi$. Moreover, for the low-energy dynamics considered here, the two-body interaction reduces approximately to $\tilde{U}(|\vec{k}|) \approx \tilde{U}(|\vec{\Delta|})$ [where 
$\Delta = \vec{k}_L-\vec{k}_R = \pi (1,1,1)$],  
resulting in the locality of the bilinears in the quartic interaction terms: using the expressions (\ref{spinors}) in the interacting non-relativistic Hamiltonian, one gets 
\beq 
\tilde{U}(|\vec{\Delta}|) \int d^4 x  
(\bar{\psi} \psi)(\bar{\psi} \psi) \,. 
\label{scalar2}
\eeq
Although not strictly renormalizable, the term (\ref{scalar2}) induces an effective interaction term (with coupling parameter having dimensions [mass]$^{-2}$, similarly to the interacting term in the Nambu-Jona Lasinio model \cite{NJL}). 
The term (\ref{scalar2}) is present also when Bragg pulses are used 
to make massive the Dirac fermions. 

Obviously the interatomic term (\ref{interaz}) could even spoil the picture in terms of Dirac spinors. However, with a weak two-body potential ($U(\vec{r}) \ll t$), the same condition (\ref{trasf}) guarantees the validity of the relativistic description \cite{nota3} since this holds as far as the lifetime $\tau$ of the quasi-particle excitations is finite and 
$\Gamma =h/\tau$ smaller than their energy, which is indeed our case (the lifetime can be estimated along the lines discussed in \cite{GV} and the details will be published elsewhere).

In the $2D$ limit eqn (\ref{scalar2}) 
remains still valid and it induces a coupling between the different spinors $\psi_{1}^{2D}$ and $\psi_{2}^{2D}$ of 
the $(2+1)$ Dirac fermions; 
moreover, relaxing condition (\ref{trasf}) -- not strictly necessary in the $2D$ limit -- one obtains all the coupling terms with an even number of 
$\psi_{1}^{2D}$ $or$ $\psi_{2}^{2D}$.

To conclude, we observe that when ${\cal N}$ different fermionic species are considered the interaction term (\ref{interaz}) reads 
$\sum_{i,j} \sum_{\alpha,\beta=1}^{\cal N}\, U_{i,j}^{\alpha,\beta} \,  c_{i;\alpha}^{\dag} c_{i;\alpha} c_{j;\beta}^{\dag} c_{j;\beta}$. 
In general, the interaction intra-($\alpha$$=$$\beta$) and inter-($\alpha$$\neq$$\beta$) species are different: however, the condition 
(\ref{trasf}), necessary to have Lorentz invariant interaction terms, simply reads $\tilde{U}^{\alpha,\beta} (0)= 0$. If 
the interaction is independent on the internal degrees of freedom $\alpha,\beta$, then eqn (\ref{scalar2}) becomes
$\tilde{U}(|\vec{\Delta}|) \int d^4 x \, 
(\bar{\psi}_{\alpha} \psi^{\alpha})(\bar{\psi}_{\beta} \psi^{\beta})$.

{\em Conclusions}: We have shown that ultracold atoms in a rotating optical lattice are able to simulate $(3+1)$ Dirac fermions, with their mass generated 
by a Bragg pulse which transfers particles from a Dirac point to the other. When the two lasers of the Bragg pulse have the same frequencies, the Dirac mass is time-independent, otherwise one has a sinusoidal time-dependence of the mass. 
This property could be used to study adiabatic or quenched dynamics in the Dirac equation; with random Bragg pulses, it can be used to investigate instead diffusion and disorder in relativistic quantum mechanics.

We have also analyzed the crossover from $(3+1)$ to $(2+1)$ Dirac fermions 
which can be induced by anisotropic lattices. Finally, we have also given a 
criterion for the interatomic interactions in order to get relativistically invariant effective interaction terms. 
Interesting perspectives along this line include the possibilities: (i) to study the relativistic Hamiltonian for several 
species with general interactions (in particular with no intra-species interactions); 
(ii) to simulate the Nambu-Jona Lasinio model \cite{NJL}; (iii)
to manipulate the ultracold atomic lattice for realizing Majorana fermions;  
(iv) to study the (eventually attractive) relativistic interacting theory which we have obtained, also in the $3D-2D$ crossover.

{\em Acknowledgements:} Useful discussions and correspondence 
with D.M. Basko, L. Dell'Anna and M. M\"uller are gratefully acknowledged. 
L.L. also thanks G. Borghi, F. Ferrari Ruffino and L. Maccione 
for fruitful discussions. 
This work is supported 
by the grants INSTANS (from ESF) and 2007JHLPEZ (from MIUR).


\begin{thebibliography}{XX}
\bibitem{nota1}
The other two choices with interchanged $(k_x , k_y , k_z)$ are equivalent, 
giving the same energy levels \cite{landau9}.
\bibitem{nota3} 
To realize the condition (\ref{trasf}), it is enough to have 
$|\tilde{U}(0)| \ll t a^3$.
\end{thebibliography}
\end{document}